# Techniques on mesh generation for the brain shift simulation

Claudio Lobos[1], Marek Bucki[1], Yohan Payan[1] and Nancy Hitschfeld[2],

1 TIMC Laboratory, UMR CNRS 5225, University Jospeh Fourier, 38706 La Tronche, France, Claudio.Lobos@imag.fr,

2 Department of Computer Science, University of Chile, Blanco Encalada 2120, Santiago, Chile,

*Abstract*— **Neurosurgery interventions involve complex tracking systems because a tissue deformation takes place. The neuronavigation system relies only on pre-operative images. In order to overcome the soft tissue deformations and guarantee the accuracy of the navigation a biomechanical model can be used during surgery to simulate the deformation of the brain. Therefore, a mesh generation for an optimal real-time Finite Element Model (FEM) becomes crucial. In this work we present different alternatives from a mesh generation point of view that were evaluated to optimize the process in terms of elements quantity and quality as well as constraints of a intraoperative application and patient specific data.**

## I. INTRODUCTION

Accurate localization of the target is essential to reduce morbidity during a brain tumor removal intervention. Image guided neurosurgery nowadays faces an important issue for large skull openings, with intra-operative changes that remain largely unsolved. The deformation causes can be grouped by:
- physical changes (dura opening, gravity, loss of cerebrospinal fluid, actions of the neurosurgeon, etc) and
- physiological phenomena (swelling due to osmotic drugs, anesthetics, etc).

As a consequence of this intra-operative brain-shift, pre-operative images no longer correspond to reality. Therefore the neuro-navigation system based on those images is strongly compromised to represent the current situation.

In order to face this problem, scientists have proposed to incorporate into existing image-guided neurosurgical systems, a module to compensate for brain deformations by updating the pre-operative images and planning according to intra-operative brain shape changes. This means that a strong modeling effort must be carried out during the design of the biomechanical model of the brain. The model must also be validated against clinical data.

The main flow of our strategy can be divided into three main steps:
- The segmentation of MRI images to build a surface mesh of the brain with the tumor.
- The generation of a volume mesh optimized for real-time simulation.
- The creation of a model of the Brain Shift with Finite Elements and the updating with ultrasound images.

This paper aims to deal with the second point: the selection of a meshing technique for this particular problem. Several algorithms and applications are analyzed and contrasted.

## II. MESHING CONSTRAINTS

Various constraints and statements have arisen in the path to achieve this surgical simulation. These global ideas can be summarized as follows:
- The speed of the FEM computation depends on the number of points the system has to deal with.
- A good representation of the tumor as well as the Opening Skull Point (OSP) and the path between them is needed. This path will from now on be referred to as the Region of Interest (RoI).
- Consideration of the brain ventricles is desirable.
- The algorithm should consider as input a surface mesh with the ventricles and the tumor.
- Obtain a surface representation and element quality throughout the entire mesh.

## III. MESHING TECHNIQUES

### A. Advancing Front

This is a technique that starts with a closed surface [1,2]. All the faces that describe the surface are treated as fronts and are expanded into the volume in order to achieve a final 3D representation. The selection of points to create the new faces encourages the use of existing points. Additional process to improve the quality of the elements can be made.

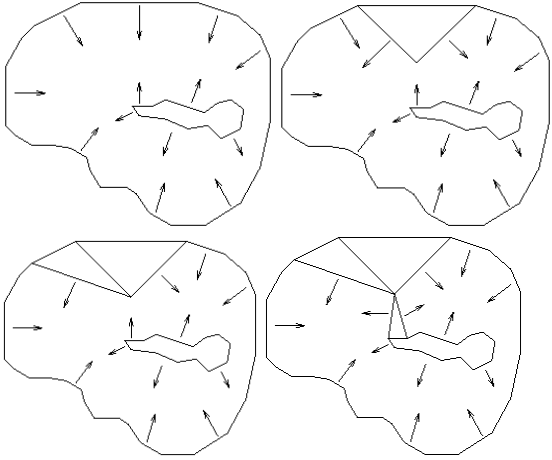

Fig. 1 The advancing front technique: top-left: the initial surface mesh with the expansions directions, top-right: one face (front) is expanded, bottom-left: another expansion using already inserted points, bottom-right: expansion of cavity inner face using inserted points.

There are two main drawbacks to this approach. The first one is that this technique is recommended when it is important to maintain the original input faces. This is not a specific constraint in our case. The second is that it would be necessary to use external libraries to produce a local or regional refinement. In our case, this is to have a refined mesh in the RoI and coarse elsewhere. However, the control of internal regions can be easily achieved because inner surface subsets are also considered to expand. All this is shown in figure 1.

*B. Mesh matching*

Mesh matching is an algorithm that starts with a generic volume mesh and tries to match it to the specific surface [3]. The base volume is obtained from an interpolation of several sample models. To obtain a new mesh, in our case for the brain, the problem is reduced to find a transformation function that will be applied to the entire base mesh and in that manner produce the final volume mesh for the current patient. This is shown in figure 2.

The problem with this technique is that even though a good representation of the surface and quality of the mesh can be achieved, it would not contain the tumor information because its position and thus the RoI changes from one patient to another.

This technique is recommended for, and has been successfully applied to complex structures such as bones and maxillofacial models [4].

A good adaptation of this technique to our problem would be to provide tools for local element refinement and with this produce a more refined mesh in the RoI. Then a subset of these elements would be labeled as the tumor, leaving the rest of the mesh untouched.

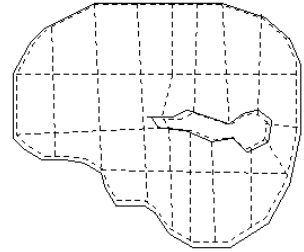

Fig. 2 The mesh matching algorithm: the segmented lines represent the base mesh that has to match the surface.

*C. Regular Octree*

The octree technique starts from the bounding box of the surface to mesh [5]. This basic cube or "octant" is split into eight new octants. Each octant is then iteratively split into eight new ones, unless it resides outside the input surface mesh, in which case it is remove from the list. The algorithm stops when a predefined maximum level of iterations is reached or when a condition of surface approximation is satisfied. This is shown in figure 3.

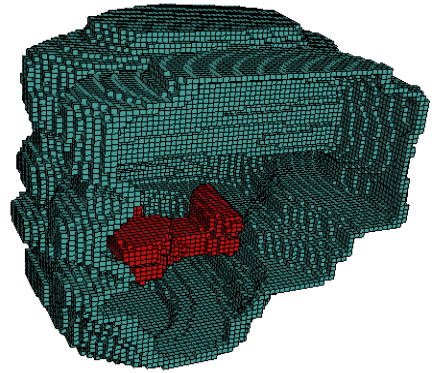

Fig. 3 Regular Octree: a regular mesh that contains hexahedral elements. Only elements that intersect the surface or the cavity are shown to better appreciate the last one. A real output mesh consider all the hexahedra that where erased this time and all of them have the same size.

The octree by itself does not consider a surface approximation algorithm once the split process is done.

Claudio Lobos, Marek Bucki, Yohan Payan and Nancy Hitschfeld.

Therefore it has to be combined with other techniques in order to produce a final mesh that represents well the surface. Two main approaches are considered:

- Marching cubes [6]: this algorithm crops the cubes that lie within the surface and produce, in most cases, tetrahedrons.
- Surface projection: this technique projects the points of those elements that intersect the surface, onto it. The main problem is that this can produce degenerated elements unless a minimal displacement is needed.

The problem with a regular octree mesh is that it counts with a high number of points even in regions where no displacement is expected [7]. Therefore, a non-optimal mesh would be the input for the FEM producing unnecessary time consumption for the entire simulation.

*D. Delaunay*

A Delaunay triangulation or Delone triangularization for a set P of points in the plane is a triangulation DT(P) such that no point in P is inside the circumcircle of any triangle in DT(P). Delaunay triangulations maximize the minimum angle of all the angles of the triangles in the triangulation. The triangulation was invented by Delaunay in 1934 [8].

To evaluate this technique we used TetGen [9]. This is an open-source application that generates tetrahedral meshes and Delaunay tetrahedralizations. The tetrahedral meshes are suitable for finite element and finite volume methods. TetGen can be used either as an executable program or as a library for integrating into other applications. A mesh generated by TetGen is shown in figure 4.

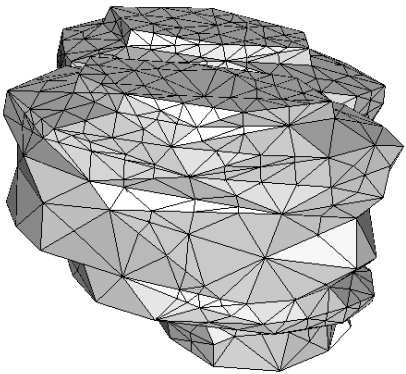

Fig. 4 One output mesh generated by TetGen.

When used as a stand alone program, a surface mesh must be provided as well as some information to identify regions and cavities. The output is a constrained Delaunay volume mesh that contains quality tetrahedral elements.

Because only one file that describes the surface (and inner regions) can be used as input, this mesh must contain the RoI within itself, i.e. several faces have to be joined in order to produce a closed region that will count with a higher level of refinement.

It is possible to not constrain the volume in the zones outside the RoI. In this case TetGen will produce elements as large as possible in those regions, while satisfying the Delaunay property as well as surface representation.

When used as a library, the situation changes. The decision of which element to split can be controlled by another class that defines some specific criteria like regional refinement. The algorithm would start from a global mesh that is not constrained by a maximum element volume. Then detect every tetrahedron that resides in the RoI and produce a refinement of those elements until a certain condition on point quantity is reached. With these modifications, all the initial constraints as defined in section II would be satisfied, producing an optimal patient-specific mesh. This approach has not been implemented.

*E. Modified Octree*

The octree structure is very convenient for refining only certain elements in the mesh. As explained before, it splits all the elements that are not completely outside the input domain. A minor conceptual modification would make a significant difference in obtaining an optimal mesh: split all the elements that are not completely outside the RoI. This creates a new category of element: an element that is outside the RoI but inside the input domain.

At this level the mesh is unsuitable for the FEM because it has some element faces that are split by one side and not by the other. In other words, there are faces that count with some points inserted on their edges and even on their surface, as shown in figure 5.

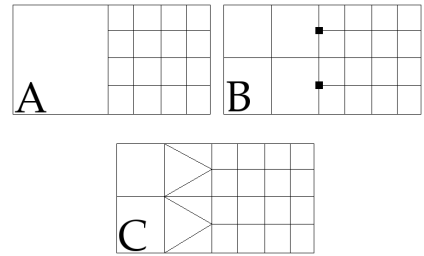

Fig. 5 Octree transitions: a) RoI output, b) 1-irregular mesh and c) mixed element valid mesh.

A mesh is said to be 1-irregular if each element face has a maximum of one point inserted on each edge and a face midpoint. The mesh produced by adding the refinement constraint at the RoI is not necessarily 1-irregular. To solve this problem, all the elements that are not 1-irregular are split until the entire mesh respects this property [10].

It is important to produce a 1-irregular mesh because throughout patterns we can split those octants by adding different type of elements (such as: tetrahedra, pyramids and prisms). This stage results in a congruent mixed element mesh that has different levels of refinement.

The last step is to achieve a representation of the surface. This is done by projecting the points of the elements that are outside the input surface, onto it. Better solutions, such as marching cubes, can be implemented with modifications in order to avoid problems with the junction of other types of elements (tetrahedra, pyramid and prism). The main motivation for changing this sub-process is that projected elements may not respect some aspect-ratio constraints as can be seen figure 6.

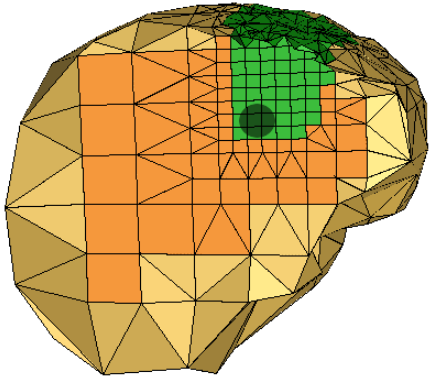

Fig. 6 One output mesh generated by the modified octree technique. The circle represents the tumor inside the RoI. Inner elements remain regular.

This modified octree has been implemented with satisfactory results. However it can still evolve further with the achievement of a better surface representation. This technique also complies with the constraints mentioned in section II.

The stop condition for the initial octree is not a certain level of surface approximation, but a global quantity of points. Once this quantity is reached, the only process that will increase the number of points is the 1-irregular process. After that, the management of the transition and the process of projection will not increase the quantity of points.

## IV. DISSCUSION

We have presented several meshing techniques that confront the problem of brain shift. In particular, we have mentioned two techniques that can respect all the initial constraints of the problem (TetGen and Modified Octree) and one technique that can be very useful when time constrain is the most important (Multi-resolution).

At this point we can summarize the future works as:
1. Adapt TetGen to the problem, using it as a library.
2. Improve the quality of the projected elements in the Modified Octree.
3. Compare Modified Octree with TetGen in terms of point quantity and element quality.

## V. ACKNOWLEDGMENT

This project has been financially supported by the ALFA IPECA project as well as FONDEF D04-I-1237 (Chile), and carried out in collaboration with Praxim-Medivision (France). The work of Nancy Hitschfeld is supported by Fondecyt project number 1061227.

Author:   Claudio Lobos
Institute: TIMC-IMAG
Street:   Faculté de médecine, pavillon Taillefer
City:     La Tronche
Country: France
Email:    Claudio.Lobos@imag.fr